\documentclass[onecolumn]{aastex701}
\usepackage{caption}
\begin{document}

\title{A Path to an All-Sky Survey with Roman\vspace{24pt}}


\author[0000-0002-6800-5778]{Jiwon Jesse Han}
\affil{Kavli Institute for Particle Astrophysics \& Cosmology, P. O. Box 2450, Stanford University, Stanford, CA 94305, USA}
\email[show]{jiwonhan@stanford.edu}

\author{Anirudh Chiti}
\affil{Kavli Institute for Particle Astrophysics \& Cosmology, P. O. Box 2450, Stanford University, Stanford, CA 94305, USA}
\email{achiti@stanford.edu}

\author[0000-0002-3839-0230]{Kai-Feng Chen}
\affil{MIT Kavli Institute for Astrophysics and Space Research, Cambridge, MA 02139, USA}
\affil{Department of Physics, Massachusetts Institute of Technology, Cambridge, MA 02139, USA}
\email{kfchen@mit.edu}

\author{Keith Bechtol}
\affil{Physics Department, 2320 Chamberlin Hall, University of Wisconsin-Madison, 1150 University Avenue Madison, WI 53706-1390}
\email{kbechtol@wisc.edu}

\author{Andrea Bellini}
\affil{Space Telescope Science Institute, 3700 San Martin Drive, Baltimore, MD 21218, USA}
\email{bellini@stsci.edu}

\author{Robert Benjamin}
\affil{Department of Astronomy, University of Wisconsin—Madison, WI, USA}
\affil{Department of Physics, University of Wisconsin—Whitewater, WI, USA}
\email{benjamir@uww.edu}

\author{Adam Bolton}
\affil{Kavli Institute for Particle Astrophysics \& Cosmology, P. O. Box 2450, Stanford University, Stanford, CA 94305, USA}
\affil{SLAC National Accelerator Laboratory, Menlo Park, CA 94025, USA}
\email{abolton@slac.stanford.edu}

\author{Ana Bonaca}
\affil{The Observatories of the Carnegie Institution for Science, Pasadena, CA 91101, USA}
\email{abonaca@carnegiescience.edu}

\author{Alex Broughton}
\affil{Kavli Institute for Particle Astrophysics \& Cosmology, P. O. Box 2450, Stanford University, Stanford, CA 94305, USA}
\affil{SLAC National Accelerator Laboratory, Menlo Park, CA 94025, USA}
\email{alex.broughton@slac.stanford.edu}

\author{Esra Bulbul}
\affil{Max Planck Institute for Extraterrestrial Physics, Giessenbachstrasse 1, 85748 Garching, Germany}
\email{ebulbul@mpe.mpg.de}

\author{Susan Clark}
\affil{Kavli Institute for Particle Astrophysics \& Cosmology, P. O. Box 2450, Stanford University, Stanford, CA 94305, USA}
\affil{Department of Physics, Stanford University, 382 Via Pueblo Mall, Stanford, CA 94305, USA}
\email{seclark1@stanford.edu}

\author{Charlie Conroy}
\affil{Center for Astrophysics $\vert$ Harvard \& Smithsonian, 60 Garden Street, Cambridge, MA 02138 USA}
\email{cconroy@cfa.harvard.edu}

\author{Suchetha Cooray}
\affil{Kavli Institute for Particle Astrophysics \& Cosmology, P. O. Box 2450, Stanford University, Stanford, CA 94305, USA}
\email{cooray@stanford.edu}

\author{John Franklin Crenshaw}
\affil{Kavli Institute for Particle Astrophysics \& Cosmology, P. O. Box 2450, Stanford University, Stanford, CA 94305, USA}
\email{jfcren@stanford.edu}

\author[0000-0002-6939-9211]{Tansu Daylan}
\affil{Department of Physics and McDonnell Center for the Space Sciences, Washington University, St. Louis, MO 63130, USA}
\email{tansu@wustl.edu}

\author{Arjun Dey}
\affil{NSF National Optical Infrared Astronomy Research Lab}
\email{arjun.dey@noirlab.edu}

\author{Alex Drlica-Wagner}
\affil{Department of Astronomy \& Astrophysics, University of Chicago, 5640 S Ellis Avenue, Chicago, IL 60637, USA}
\affil{Kavli Institute for Cosmological Physics, University of Chicago, Chicago, IL 60637, USA}
\email{kadrlica@uchicago.edu}

\author{Tim Eifler}
\affil{Department of Astronomy and Steward Observatory, University of Arizona, 933 North Cherry Avenue, Tucson, AZ 85721, USA}
\affil{Department of Physics, University of Arizona, 1118 E. Fourth Street, Tucson, AZ 85721, USA}
\email{timeifler@arizona.edu}

\author[0000-0002-6871-1752]{Kareem El-Badry}
\affil{Department of Astronomy, California Institute of Technology, 1200 E. California Blvd., Pasadena, CA 91125, USA}
\email{kelbadry@caltech.edu}

\author[0000-0001-5382-6138]{Richard M. Feder}
\affil{Berkeley Center for Cosmological Physics, University of California, Berkeley, CA 94720, USA}
\affil{Lawrence Berkeley National Laboratory, 1 Cyclotron Road, Berkeley, CA 94720, USA}
\email{rmfeder@berkeley.edu}

\author{Peter Ferguson}
\affil{DiRAC Institute and Department of Astronomy, University of Washington, 3910 15th Ave NE, Seattle, WA, 98195, USA}
\email{pferguso@uw.edu}

\author{Shenming Fu}
\affil{Kavli Institute for Particle Astrophysics \& Cosmology, P. O. Box 2450, Stanford University, Stanford, CA 94305, USA}
\email{sfu@slac.stanford.edu}

\author[0000-0001-6395-6702]{Sebastian Gomez}
\affil{Department of Astronomy, The University of Texas at Austin, 2515 Speedway, Stop C1400, Austin, TX 78712, USA}
\email{sebastian.gomez@austin.utexas.edu}

\author{Ryan Hickox}
\affil{Department of Physics \& Astronomy, Dartmouth College, Hanover, NH 03755}
\email{Ryan.C.Hickox@dartmouth.edu}

\author{Christopher Hirata}
\affil{Center for Cosmology and Astroparticle Physics, The Ohio State University, 191 West Woodruff Avenue, Columbus, Ohio 43210, USA}
\affil{Department of Physics, The Ohio State University, 191 West Woodruff Avenue, Columbus, Ohio 43210, USA}
\affil{Department of Astronomy, The Ohio State University, 140 West 18th Avenue, Columbus, Ohio 43210, USA}
\email{hirata.10@osu.edu}

\author{Easton J. Honaker}
\affil{Department of Physics and Astronomy, University of Delaware, Newark, DE 19716, USA}
\email{ehonaker@udel.edu}

\author{Xiaosheng Huang}
\affil{Department of Physics \& Astronomy, University of San Francisco, San Francisco, CA 94117}
\email{xhuang22@usfca.edu}

\author[0000-0002-4863-8842]{Alexander P. Ji}
\affil{Department of Astronomy \& Astrophysics, University of Chicago, 5640 S Ellis Avenue, Chicago, IL 60637, USA}
\affil{Kavli Institute for Cosmological Physics, University of Chicago, Chicago, IL 60637, USA}
\affil{NSF-Simons AI Institute for the Sky (SkAI), 172 E. Chestnut St., Chicago, IL 60611, USA}
\email{alexji@uchicago.edu}

\author{Michael Liu}
\affil{Institute for Astronomy, University of Hawai'i, 2680 Woodlawn Drive, Honolulu HI 96822}
\email{liumicha@hawaii.edu}

\author[0000-0001-7494-5910]{Kevin A. McKinnon}
\affil{David A. Dunlap Department of Astronomy \& Astrophysics, University of Toronto, 50 St George Street, Toronto ON M5S 3H4, Canada}
\email{kevinmckinnon95@gmail.com}

\author[0000-0001-6331-112X]{Geoffrey Mo}
\affil{Division of Physics, Mathematics and Astronomy, California Institute of Technology, Pasadena, CA 91125, USA}
\affil{The Observatories of the Carnegie Institution for Science, Pasadena, CA 91101, USA}
\email{gmo@carnegiescience.edu}

\author{Burcin Mutlu-Pakdil}
\affil{Department of Physics \& Astronomy, Dartmouth College, Hanover, NH 03755}
\email{Burcin.Mutlu-Pakdil@dartmouth.edu}

\author[0000-0003-0872-7098]{Adrian M. Price-Whelan}
\affil{Center for Computational Astrophysics, Flatiron Institute, 162 Fifth Ave, New York, NY 10010, USA}
\email{aprice-whelan@flatironinstitute.org}

\author[0000-0002-1445-4877]{Alessandro Savino}
\affil{Department of Astronomy, University of California, Berkeley, Berkeley, CA 94720, USA}
\email{asavino@berkeley.edu}

\author{David J. Schlegel}
\affil{Lawrence Berkeley National Laboratory, 1 Cyclotron Road, Berkeley, CA 94720, USA}
\email{DJSchlegel@lbl.gov}

\author{Nora Shipp}
\affil{DiRAC Institute and Department of Astronomy, University of Washington, 3910 15th Ave NE, Seattle, WA, 98195, USA}
\email{nshipp@uw.edu}

\author{Jay Strader}
\affil{Center for Data Intensive and Time Domain Astronomy, Department of Physics and Astronomy,\\ Michigan State University, East Lansing, MI 48824, USA}
\email{straderj@msu.edu}

\author{Federica Tarsitano}
\affil{Institute for Particle Physics and Astrophysics, Dept. of Physics, ETH Zurich, Wolfgang-Pauli-Strasse 27, 8093 Zurich, Switzerland}
\email{federica.tarsitano@phys.ethz.ch}

\author[0000-0001-7928-1973]{Adrien C.R. Thob}
\affil{Department of Physics \& Astronomy, University of Pennsylvania, Philadelphia, PA 19104, USA}
\email{athob@sas.upenn.edu}

\author{Kim-Vy Tran}
\affil{Center for Astrophysics $\vert$ Harvard \& Smithsonian, 60 Garden Street, Cambridge, MA 02138 USA}
\email{kim-vy.tran@cfa.harvard.edu}

\author[0000-0001-7827-7825]{Roeland P. van der Marel}
\affil{Space Telescope Science Institute, 3700 San Martin Drive, Baltimore, MD 21218, USA}
\affil{Center for Astrophysical Sciences, The William H. Miller III Department of Physics \& Astronomy, Johns Hopkins University, Baltimore, MD 21218, USA}
\email{marel@stsci.edu}

\author{Feige Wang}
\affil{Department of Astronomy, University of Michigan, Ann Arbor, MI 48109, USA}
\email{fgwang@umich.edu}

\author{Risa Wechsler}
\affil{Kavli Institute for Particle Astrophysics \& Cosmology, P. O. Box 2450, Stanford University, Stanford, CA 94305, USA}
\affil{SLAC National Accelerator Laboratory, Menlo Park, CA 94025, USA}
\affil{Department of Physics, Stanford University, 382 Via Pueblo Mall, Stanford, CA 94305, USA}
\email{rwechsler@stanford.edu}

\author[0000-0002-6442-6030]{Daniel~R.~ Weisz}
\affil{Department of Astronomy, University of California, Berkeley, CA 94720-3411, USA}
\email{dan.weisz@berkeley.edu}

\author{Dennis Zaritsky}
\affil{Department of Astronomy and Steward Observatory, University of Arizona, 933 North Cherry Avenue, Tucson, AZ 85721, USA}
\email{dfz@arizona.edu}

\author{Tianqing Zhang}
\affil{Department of Physics and Astronomy and PITT PACC, University of Pittsburgh, Pittsburgh, PA 15260}
\email{tq.zhang@pitt.edu}

\author[orcid=0000-0003-3768-7515]{Shreya Anand}
\affil{Kavli Institute for Particle Astrophysics \& Cosmology, P. O. Box 2450, Stanford University, Stanford, CA 94305, USA}
\email{sanand08@stanford.edu}

\author{Amirnezam Amiri}
\affil{School of Astronomy, Institute for research in fundamental sciences
(IPM), Tehran, P.O. Box 19395-5531, Iran}
\affil{Department of Physics, University of Arkansas, 226 Physics Build-
ing, 825 West Dickson Street, Fayetteville, AR 72701, USA}
\email{amirnezamamiri@gmail.com}

\author{Abhijeet Anand}
\affil{Inter-University Centre for Astronomy \& Astrophysics, Post Bag 04, Pune, India 411007}
\affil{Lawrence Berkeley National Laboratory, 1 Cyclotron Road, Berkeley, CA 94720, USA}
\email{abhijeet.anand@iucaa.in}

\author{Matthew L.~N.~ Ashby}
\affil{Center for Astrophysics $\vert$ Harvard \& Smithsonian, 60 Garden Street, Cambridge, MA 02138 USA}
\email{mashby@cfa.harvard.edu}

\author[0000-0001-9416-5874]{Finian Ashmead}
\affil{Department of Physics and Astronomy and PITT PACC, University of Pittsburgh, Pittsburgh, PA 15260}
\affil{Pittsburgh Particle Physics, Astrophysics, and Cosmology Center (PITT PACC), University of Pittsburgh, Pittsburgh, PA 15260, USA}
\email{finianashmead@pitt.edu}

\author{Leandro Beraldo e Silva}
\affil{Observatório Nacional, Rio de Janeiro - RJ, 20921-400, Brasil}
\email{lberaldoesilva@on.br}

\author{Aliza Beverage}
\affil{The Observatories of the Carnegie Institution for Science, Pasadena, CA 91101, USA}
\email{}

\author{Michael R. Blanton}
\affil{The Observatories of the Carnegie Institution for Science, Pasadena, CA 91101, USA}
\email{mblanton@carnegiescience.edu}

\author[0000-0002-4462-2341]{Warren R. Brown}
\affil{Center for Astrophysics $\vert$ Harvard \& Smithsonian, 60 Garden Street, Cambridge, MA 02138 USA}
\email{wbrown@cfa.harvard.edu}

\author{Anthony G.A. Brown}
\affil{Leiden Observatory, Leiden University, Einsteinweg 55, 2333
CC Leiden, The Netherlands}
\email{brown@strw.leidenuniv.nl}

\author{Priyanka Chakraborty}
\affil{Department of Physics, University of Arkansas, 825 W Dickson st., Fayetteville, AR 72701, USA}
\affil{Center for Astrophysics $\vert$ Harvard \& Smithsonian, 60 Garden Street, Cambridge, MA 02138 USA}
\email{pchakraborty@uark.edu}

\author{Yi-Kuan Chiang}
\affiliation{Academia Sinica Institute of Astronomy and Astrophysics (ASIAA), Taipei 106319, Taiwan}
\email{ykchiang@asiaa.sinica.edu.tw}

\author{Jose M. Diego}
\affil{Instituto de Fisica de Cantabria (CSIC-UC)}
\email{jdiego@ifca.unican.es}

\author{Denis Erkal}
\affiliation{School of Mathematics and Physics, University of Surrey, Guildford, GU2 7XH, UK}
\email{d.erkal@surrey.ac.uk}

\author{Simone Ferraro}
\affil{Lawrence Berkeley National Laboratory, 1 Cyclotron Road, Berkeley, CA 94720, USA}
\affil{Berkeley Center for Cosmological Physics, University of California, Berkeley, CA 94720, USA}
\email{sferraro@lbl.gov}

\author{Llu\'is Galbany}
\affil{Institute of Space Sciences (ICE-CSIC), Campus UAB, 
Carrer de Can Magrans, s/n, E-08193 Barcelona, Spain.}
\affil{Institut d'Estudis Espacials de Catalunya (IEEC), 08860 
Castelldefels (Barcelona), Spain}
\email{l.g@csic.es}

\author{Marla Geha}
\affil{Department of Astronomy, Yale University, New Haven, CT 06520, USA}
\email{marla.geha@yale.edu}

\author{Oleg~Y. Gnedin}
\affil{Department of Astronomy, University of Michigan, Ann Arbor, MI 48109, USA}
\email{ognedin@umich.edu}

\author{Lars Hernquist}
\affil{Center for Astrophysics $\vert$ Harvard \& Smithsonian, 60 Garden Street, Cambridge, MA 02138 USA}
\email{lhernquist@cfa.harvard.edu}

\author{Jason A. S. Hunt}
\affil{School of Mathematics \& Physics, University of Surrey, Stag Hill, Guildford, GU2 7XH, UK}
\email{j.a.hunt@surrey.ac.uk}

\author{Valentin D. Ivanov}
\affil{European Southern Observatory, Karl-Schwarszchild-Str. 2, D-85748, Garching bei M\"unchen, Germany}
\email{vivanov@eso.org}

\author{Venu Kalari}
\affil{Gemini Observatory/NSF’s NOIRLab, Casilla 603, La Serena, Chile}
\email{venu.kalari@noirlab.edu}

\author[orcid=0000-0002-3204-1742]{Nitya Kallivayalil}
\affiliation{Department of Astronomy, University of Virginia, 530 McCormick Road, Charlottesville, VA 22904, USA}
\email{njk3r@virginia.edu}

\author{András Kovács}
\affil{MTA–CSFK Lendület “Momentum” Large-Scale Structure Research Group, Konkoly Thege Miklós út 15-17, H-1121 Budapest, Hungary}
\affil{Konkoly Observatory, HUN-REN Research Centre for Astronomy and Earth Sciences, Konkoly Thege Miklós út 15-17, H-1121 Budapest, Hungary}
\email{kovacs.andras@csfk.org}

\author{Kyle Kremer}
\affil{Department of Astronomy \& Astrophysics, University of California, San Diego; La Jolla, CA 92093, USA}
\email{kykremer@ucsd.edu}

\author{Ting-Wen Lan}
\affil{Graduate Institute of Astrophysics and Department of Physics, National Taiwan University, No. 1, Sec. 4, Roosevelt Rd., Taipei 10617, Taiwan}
\email{twlan@ntu.edu.tw}

\author{Denis Leahy}
\affil{Department of Physics and Astronomy, University of Calgary, Calgary, AB, Canada T2N 1N4}
\email{leahy@ucalgary.ca}

\author[0000-0001-9592-4190]{Jiaxuan Li}
\affil{Department of Astrophysical Sciences, 4 Ivy Lane, Princeton University, Princeton, NJ 08540}
\email{jiaxuanl@princeton.edu}

\author{Ivan Minchev}
\affil{Leibniz Institute for Astrophysics Potsdam (AIP), Potsdam, Germany}
\email{iminchev@aip.de}

\author{GyuChul Myung}
\affil{Institute of Astronomy, University of Cambridge}
\email{gm564@cam.ac.uk}

\author{Ethan O. Nadler}
\affil{Department of Astronomy \& Astrophysics, University of California, San Diego, La Jolla, CA 92093, USA}
\email{enadler@ucsd.edu}

\author{Joan R. Najita}
\affil{NSF National Optical Infrared Astronomy Research Lab}
\email{joan.najita@noirlab.edu}

\author{Melissa K. Ness}
\affil{Research School of Astronomy \& Astrophysics, Australian National University, Canberra ACT 2611, Australia}
\email{melissa.ness@anu.edu.au}

\author{Jacob Nibauer}
\affil{Department of Astrophysical Sciences, 4 Ivy Lane, Princeton University, Princeton, NJ 08540}
\email{jnibauer@princeton.edu}

\author{Fabio Pacucci}
\affil{Center for Astrophysics $\vert$ Harvard \& Smithsonian, 60 Garden Street, Cambridge, MA 02138 USA}
\email{fabio.pacucci@cfa.harvard.edu}

\author{David Parkinson}
\affil{Korea Astronomy and Space Science Institute, 776 Daedeok-daero, Yuseong-gu, Daejeon 34055, Republic of Korea}
\email{davidparkinson@kasi.re.kr}

\author[0000-0002-9820-1219]{Ekta Patel}
\affil{Department of Astrophysics and Planetary Sciences, Villanova University,  800 E. Lancaster Ave, Villanova, PA 19085, USA}
\email{ekta.patel@villanova.edu}

\author{R. Michael Rich}
\affil{Department of Physics and Astronomy, UCLA, PAB 430 Portola Plaza, LA, CA 90095-1547}
\email{rmr@astro.ucla.edu}

\author{Marina Ricci}
\affiliation{Université Paris Cité, CNRS, Astroparticule et Cosmologie, F-75013 Paris, France}
\email{ricci@apc.in2p3.frl}

\author{Graziano Rossi}
\affil{Department of Physics and Astronomy, Sejong University, Seoul, 143-747, Korea}
\email{graziano@sejong.ac.kr}

\author{Nikolina Sarcevic}
\affil{Department of Physics, Duke University, Science Dr, Durham, NC 27710, USA}
\email{nikolina.sarcevic@duke.edu}

\author{Arnab Sarkar}
\affil{Department of Physics, University of Arkansas, 825 W Dickson st., Fayetteville, AR 72701, USA}
\affil{MIT Kavli Institute for Astrophysics and Space Research, Cambridge, MA 02139, USA}
\email{arnabs@uark.edu}

\author[0000-0002-6561-9002]{Andrew K. Saydjari}
\affil{Department of Astrophysical Sciences, 4 Ivy Lane, Princeton University, Princeton, NJ 08540}
\email{aksaydjari@gmail.com}

\author{Arman Shafieloo}
\affil{Korea Astronomy and Space Science Institute, 776 Daedeok-daero, Yuseong-gu, Daejeon 34055, Republic of Korea}
\email{shafieloo@kasi.re.kr}

\author{Zachary Slepian}
\affil{Department of Astronomy, University of Florida}
\email{zslepian@ufl.edu}

\author[0000-0001-8368-0221]{Sangmo Tony Sohn}
\affil{Space Telescope Science Institute, 3700 San Martin Drive, Baltimore, MD 21218, USA}
\email{tsohn@stsci.edu}

\author{David N. Spergel}
\affil{Flatiron Institute, 162 Fifth Avenue, NY NY 10011}
\email{dspergel@flatironinstitute.org}

\author{R\'obert Szab\'o}
\affil{Konkoly Observatory, HUN-REN Research Centre for Astronomy and Earth Sciences, Konkoly Thege Miklós út 15-17, H-1121 Budapest, Hungary}
\email{szabo.robert@csfk.org}

\author{Christina C. Williams}
\affil{NSF National Optical Infrared Astronomy Research Lab}
\email{christina.williams@noirlab.edu}

\author{John F. Wu}
\affil{Space Telescope Science Institute, 3700 San Martin Drive, Baltimore, MD 21218, USA}
\email{jowu@stsci.edu}

\collaboration{all}{NANCY Working Group}


\thispagestyle{empty}
\clearpage

\begin{abstract}

A deep, space-based, all-sky near-infrared survey carried out with the Nancy Grace Roman Space Telescope would constitute a foundational astronomical infrastructure for decades to come. In this white paper, we present a concrete and feasible path to imaging the entire sky at $\sim0.1''$ resolution, beginning with high-impact fields in Cycle 1 and scaling to ultra-wide coverage within the nominal mission. This first-epoch survey will reach $\mathrm{H}\sim25.5$ AB mag (5$\sigma$) and maximize synergies with contemporaneous observatories, while preserving substantial time for other ambitious Roman programs. We outline representative scheduling scenarios and an example Cycle 1 program that triples early Roman–LSST overlap and delivers high-value community data products such as LSST forced photometry, joint \textit{Gaia}–Roman astrometry, and catalogs of Galactic substructure, stong lenses, and other rare systems. The Cycle 1 program will lay the foundation for an eventual all-sky survey, while also delivering high-impact early science. We invite broad community participation in shaping and carrying out both the initial program and the long-term vision of an all-sky Roman survey.

\end{abstract}

\section{Motivation and Vision}

A deep, all-sky near-infrared image with $\sim0.1"$ angular resolution and space-based image quality would constitute a foundational dataset for astronomy. Just as photographic sky surveys \citep{DSS20}, digitized optical \citep{york00} and infrared surveys \citep{skrutskie06, wright10}, and all-sky astrometric catalogs \citep{gaia16} have defined successive eras of discovery, a complete $\sim0.1"$ resolution near-infrared view of the sky would establish a lasting reference against which a broad range of astrophysical phenomena can be studied. Such a dataset would enable precise source identification and deblending across the entire sky, robust star–galaxy separation at faint magnitudes, and uniform imaging of galaxies and resolved stellar systems that would inform virtually all future ground and space-based astronomical surveys. In this sense, an all-sky Roman survey is not simply a large observing program, but a piece of shared scientific infrastructure that would underpin both discovery-driven and targeted science for decades. Once completed, an all-sky survey reaching $H\sim 25.5$ AB mag would deliver the largest resolved photometric galaxy catalog to date.

The scientific case for executing this survey, in part or in full, within the nominal five-year mission is particularly compelling. Surveying a large fraction of the sky early would enable discovery-driven science and ensure opportunities for targeted follow-up throughout the remainder of the mission. Even with a single deep near-infrared band, a wide-area Roman survey would deliver a lasting scientific legacy: all-sky space-based reference images for transients; precise galaxy shape measurements for cosmology; star---galaxy separation for faint stellar populations and probes of dark matter substructure; a powerful discovery engine for rare objects; detections of red giant branch stars within the entire local volume. In addition, combined with other major concurrent data from \textit{Euclid} and the Rubin Observatory, this survey will provide the deepest all-sky optical---near-infrared colors that will enable robust identifications of  obscured AGN, compact red sources, and high-redshift quasar populations while enhancing photometric-redshift constraints. Ultimately, this survey would stand as a definitive demonstration of Roman’s unique capabilities: wide field of view, rapid slew-and-settle, and near-IR sensitivity combined in a single transformative program. 

The timeliness of an early all-sky survey is further strengthened by Roman’s overlap with several major facilities whose concurrent observations can multiply the scientific return of Roman itself. Notably, the Rubin Observatory will be conducting the Legacy Survey of Space and Time (LSST) contemporaneously with Roman operations. As the two highest-priority ground and space-based facilities identified by the Astro2010 decadal survey, Rubin and Roman together represent a cornerstone of US leadership in astrophysics. Maximizing the scientific synergy between these missions is therefore of strategic importance. For example, LSST provides deep, multi-band, multi-epoch optical photometry that would be prohibitively expensive to obtain with Roman alone, while Roman’s space-based imaging can deliver sharp, stable point-spread functions that dramatically improve source classification and deblending for LSST, especially for red sources and $\sim50\%$ of the LSST Wide-Fast-Deep footprint that is not covered by \textit{Euclid}. A Roman survey that covers the full LSST footprint within the nominal five-year mission will deliver early, high-impact science across astrophysics and cosmology, while also establishing a deep, resolved near-infrared reference image for LSST’s time-domain discovery of transient phenomena. Beyond its temporal synergy with LSST, this survey would also deliver unique time-domain science in its own right. When compared to prior near-infrared surveys (e.g., 2MASS, VHS, VVV/X, UKIDSS), even a single-epoch Roman imaging would enable population-averaged variability studies over multi-year baselines for uncrowded sources.

An additional strategically important synergy is with JWST, NASA’s current flagship observatory. JWST’s nominal mission lifetime of roughly 5–10 years implies substantial temporal overlap with Roman’s nominal five-year mission. While Roman’s wide field of view enables discovery across vast areas of sky, JWST is unmatched in angular resolution and sensitivity, making it uniquely suited for detailed follow-up of rare systems such as massive quiescent galaxies at $z \gtrsim 2$, compact starbursts and other short-lived phases of galaxy evolution, dual active galactic nuclei, and highly constraining (``jackpot'') strong gravitational lenses, as well as faint brown dwarfs and higher-order stellar systems. Conducting wide-area searches early in the Roman mission therefore maximizes the opportunity for timely JWST follow-up of the rarest and most significant discoveries.

Early execution of an all-sky Roman survey is also critical for long proper motion baselines. The transformative impact of stellar proper motions has been clearly demonstrated by \textit{Gaia}, and Roman has the potential to extend high-precision proper motion measurements to sources significantly fainter than the \textit{Gaia} limit. However, unlike \textit{Gaia} or LSST—both of which achieve exquisite astrometric precision through upwards of a hundred repeat observations—Roman cannot execute a similar large-$N$-visit strategy across tens of thousands of square degrees. Instead, Roman must achieve its astrometric leverage primarily using only a few epochs separated by long temporal baselines. Given a $>5$-year baseline, Roman can outperform LSST proper motions in crowded or extinguished regions, especially for red sources such as giant branch stars and brown dwarfs. An early all-sky reference epoch is therefore essential: it guarantees that all future Roman observations automatically provide a second epoch, enabling proper-motion measurements across the sky. Delaying such a survey would forfeit these temporal synergies and irreversibly weaken Roman’s astrometric leverage relative to other flagship observatories. Importantly, even the first epoch alone delivers immediate scientific value. When combined with \textit{Gaia}, it can substantially improve proper motion measurements, reaching factors of 6 improvement over \textit{Gaia} DR4 for sources toward $G \sim 21$. At even fainter magnitudes ($G > 21$), the Roman first epoch can be combined with \textit{Euclid} and LSST to enable proper-motion measurements beyond the nominal \textit{Gaia} limit.

Together, these considerations motivate a clear long-term vision: to use Roman’s unique wide-field imaging capabilities to deliver a complete, high-resolution near-infrared map of the sky, executed in a manner that maximizes contemporaneous overlap with other major facilities and establishes a lasting foundation for astronomical discoveries. A comprehensive discussion of the science enabled by Roman and LSST has been presented elsewhere \cite[e.g.,][]{eifler21, gezari22, eifler24}; rather than compiling an exhaustive list of applications, this white paper focuses on articulating a concrete and executable path toward imaging the entire sky with Roman under realistic mission constraints.

The concept of an all-sky Roman survey was first presented in \citet{han23c}, leading to the formation of the Next-generation All-sky Near-infrared Community surveY (NANCY) working group. With Roman’s launch scheduled for Fall 2026 and the Core Community Surveys (CCS) now established, we outline a strategy focused on what could be accomplished within the nominal five-year mission, including a possible Cycle 1 program that would begin this effort. For readers less familiar with the current mission structure, approximately 389 days are allocated to General Astrophysics Surveys (GAS), of which 50 days are available for proposal in Cycle 1, corresponding to the first two years of observations (2027-2029). Because this dataset is envisioned as shared scientific infrastructure for a large and diverse community, its success will depend on sustained participation across many fields and career stages. We view the proposed Cycle 1 program as the first step in this collective effort, and invite the community to help shape and execute it. Interested readers can provide input through our online form 
(\href{https://forms.gle/PmAFLYy1noM28z7RA}{link here}).

\section{Survey Design Principles}

Given the ultimate goal of imaging the entire sky in the near infrared, our survey design prioritizes survey speed and long-term legacy value. For high Galactic latitude ($|b|>20^\circ$) fields, a natural choice for the primary filter is F158 (H band). F158 is the reddest Roman filter that maintains high sensitivity and has been selected as the wide-tier filter for the High Latitude Wide Area Survey (HLWAS), which will image approximately $5000\,\mathrm{deg}^2$. An H-band extension to HLWAS therefore provides a natural and efficient pathway to covering the remainder of the sky within the nominal mission. Meanwhile, the F146 filter offers the highest sensitivity—effectively combining Y, J, and H bands and reaching roughly one magnitude deeper than F158 for a fixed exposure time—but is expected to suffer from chromatic PSF effects that limit its utility for precise shape measurements. This limitation is not critical at low Galactic latitudes ($|b|<20^\circ$), where the vast majority of sources are point-like. In these regions, F146 enables a survey speed 2-3 times faster than F158, allowing rapid coverage of the Galactic plane and bulge. For example, the low latitude fields can be covered in a total of 67 days using the wide F146 filter and a 2-point dither, which is consistent with the Galactic Plane Survey's wide tier dither strategy. Given the distinct science drivers and optimal observing strategies for the Galactic plane versus the rest of the sky, we focus in this white paper on articulating a clear and executable strategy for the high latitude fields using the F158 filter, while noting that efficient coverage of the low latitude fields can be achieved in parallel using the wide F146 filter.

Given the choice of F158, we optimize for the fastest survey speed that preserves weak lensing capabilities, i.e., maintaining the total number of galaxy shape measurements for a fixed observing time. As demonstrated by the HLWAS definition team \citep{rotac25}, the total number of galaxy shapes is nearly constant over a broad range of exposure times, with values between $91\,\mathrm{s}$ and $149\,\mathrm{s}$ remaining within $\sim1\%$ of the maximum. Because Roman exposure times are discretized and must be selected from the available MultiAccum (MA) table, we adopt $\mathrm{IM\_85\_7}$, corresponding to an $85\,\mathrm{s}$ exposure. This choice preserves nearly the same total number of galaxy shape measurements as the $107\,\mathrm{s}$ exposures adopted for the HLWAS wide (and medium) tiers, while increasing the survey speed by approximately $25\%$. 

Assuming the same tiling and dither strategy as HLWAS, this corresponds to a first-epoch survey speed of approximately $81\,\mathrm{deg}^2\,\mathrm{day}^{-1}$. We adopt the HLWAS 3-point dither (LINEGAP3\_3) as the baseline configuration to ensure homogeneity with the existing wide tier, and to fully eliminate chip gaps in the first epoch. By contrast, a 2-point dither would leave approximately $1.5\%$ of the area uncovered. If early HLWAS calibration efforts ultimately demonstrate that a 2-point strategy is sufficient for wide-area mapping, it can be implemented in subsequent observations. However, should a 2-point dither prove insufficient, adopting a 3-point dither from the outset ensures robust sky coverage. For the adopted configuration of $85\,\mathrm{s}$ exposures and a 3-point dither, survey overheads amount to approximately $24\%$, which we consider sufficiently low. For typical zodiacal background levels, three $85$\,s exposures reach a $5\sigma$ point-source depth of $\mathrm{H} = 25.5$\,AB mag ($25.7$\,mag in the absence of zodiacal background). In the initial pass, $55\%$ of the survey area will receive all three exposures, achieving the full $\mathrm{H} = 25.5$ depth; $37\%$ will receive two exposures, reaching $\mathrm{H}= 25.3$; and $4\%$ will receive a single exposure, reaching $\mathrm{H} =24.9$. 

As a concrete scientific application of this depth, consider the problem of mapping resolved stellar populations. An $\mathrm{H} = 25.5$ survey reaches metal-poor main-sequence turnoff stars (MSTO) out to ${\sim}200$\,kpc. Reaching the MSTO adds invaluable information beyond the red giant branch alone, as the MSTO is sensitive to stellar ages and dramatically increases the number of available stellar tracers relative to the giant branch. While such stars will be detected by LSST ($g \lesssim 27.4$), they are too faint for reliable LSST proper motion measurements or color-based star--galaxy separation, making morphological star--galaxy separation essential.  Roman’s H-band imaging at $\sim 0.135''$ resolution can provide the necessary separation, delivering a volume-complete census of dwarf galaxies, star clusters, and stellar streams throughout the Galaxy’s virial radius. Beyond the Galaxy, an $\mathrm{H} = 25.5$ survey will detect red giant branch stars out to several Mpc and reach the tip of the red giant branch at distances of $\sim 10$--$15$\,Mpc, encompassing the full Local Volume. In regions without LSST coverage, Roman H-band data can be combined with comparably deep \textit{Euclid} VIS imaging ($25.5$\,mag) to construct broad optical--near-infrared color–magnitude diagrams. \textit{Euclid}’s near-infrared imaging is both shallower ($24$\,mag) and coarser resolution (${\sim}0.3''$), making Roman uniquely powerful at detecting the faintest and reddest stellar populations in nearby dwarf galaxies and stellar streams, which are powerful probes of galaxy formation and dark matter physics \citep[e.g.,][]{drlica-wagner19}.

Another powerful application of an $\mathrm{H} = 25.5$ survey is the identification of high-redshift quasars in combination with LSST photometry. At this depth, a $10{,}000$\,deg$^2$ $H$-band program within the LSST footprint would detect more than ten million quasars spanning $0 < z < 5$, assuming the luminosity function of \citet{kulkarni19}. In addition, this dataset would open a new discovery space by enabling the identification of tens of thousands of $z > 5$ quasar candidates for subsequent ground-based spectroscopic confirmation. Such a transformative sample would address central questions in quasar and galaxy evolution, including the growth history of supermassive black holes, the evolution of the quasar luminosity function, quasar clustering and halo occupation, and the connection between quasars and large-scale structure from the epoch of reionization to the present day.

\section{A Path to the All-Sky First Epoch}

There is a direct tradeoff between the total sky area imaged during the nominal mission and the fraction of GAS time allocated to this program. We therefore outline three representative scenarios that bracket a reasonable lower and upper bound on this tradeoff.

There are three main regions of sky lacking planned Roman coverage: the $|b|>20^{\circ}$ LSST footprint ($9643\,\mathrm{deg}^2$, requiring 119 days in F158), the $|b|>20^{\circ}$ non-LSST footprint ($12502\,\mathrm{deg}^2$, requiring 154 days in F158), and the low-Galactic latitude footprint ($|b|<20^{\circ}$; $13417\,\mathrm{deg}^2$, requiring 67 days in F146). We summarize these three regions in Table \ref{tab:sky_segments}.
 
In the conservative scenario, we prioritize coverage of the LSST extragalactic footprint alone, requiring 119 days, or approximately $30\%$ of the available GAS time. For example, this allocation could be distributed as 14 days in Cycle 1, 60 days in Cycle 3, and 45 days in Cycle 4. In the intermediate scenario, we add the Galactic plane, increasing the allocation to approximately $48\%$ of GAS time. In the most ambitious scenario, we extend coverage to all three regions, amounting to approximately $87\%$ of GAS time and enabling imaging of the entire sky within the nominal mission.

These scenarios define a continuum of options rather than discrete choices. Intermediate allocations between these benchmarks are straightforward to construct, allowing the community to balance survey scope against time allocation within the nominal mission. Moreover, these estimates assume that no additional wide-field imaging programs are executed; in practice, complementary surveys will almost certainly contribute additional area. Any remaining footprint not completed during the nominal mission can be prioritized early in the extended mission.

\begin{table*}
\centering
\caption{Three major regions of the sky to complete within the General Astrophysics Surveys (GAS) to enable an all-sky Roman Survey. Within the non-LSST $|b|>20^\circ$ footprint, 7044\,deg$^2$ (56\%) are in the \textit{Euclid} Wide Survey.}
\label{tab:sky_segments}
\begin{tabular}{lccc}
\hline
 & LSST $|b|>20^{\circ}$ 
 & $|b|<20^{\circ}$ 
 & non-LSST $|b|>20^{\circ}$ \\
\hline
Area (deg$^2$) 
& 9,643 
& 13,417 
& 12,502 \\

Filter 
& F158 
& F146 
& F158 \\

Days Required 
& 119 
& 67 
& 154 \\

Fraction of GAS Time 
& $\sim 30\%$ 
& $\sim 17\%$ 
& $\sim 40\%$ \\
\hline
\end{tabular}
\end{table*}






\section{One Possible Cycle 1 Strategy}

Cycle 1, encompassing the first two years of the mission (2027-2029), offers limited time for new programs due to commissioning activities, coronagraph demonstrations, and scheduled Core Community Surveys. Approximately 50 days are available for General Astrophysics Surveys, with an anticipated selection of 4--6 teams. Within these constraints, we outline one possible Cycle 1 program that would establish the foundation for a future all-sky survey, deliver high-impact early science, and demonstrate the unique capabilities of Roman’s wide-field survey instrument.

\subsection{Footprint}

\begin{figure}[ht]
    \centering
    \includegraphics[width=\linewidth]{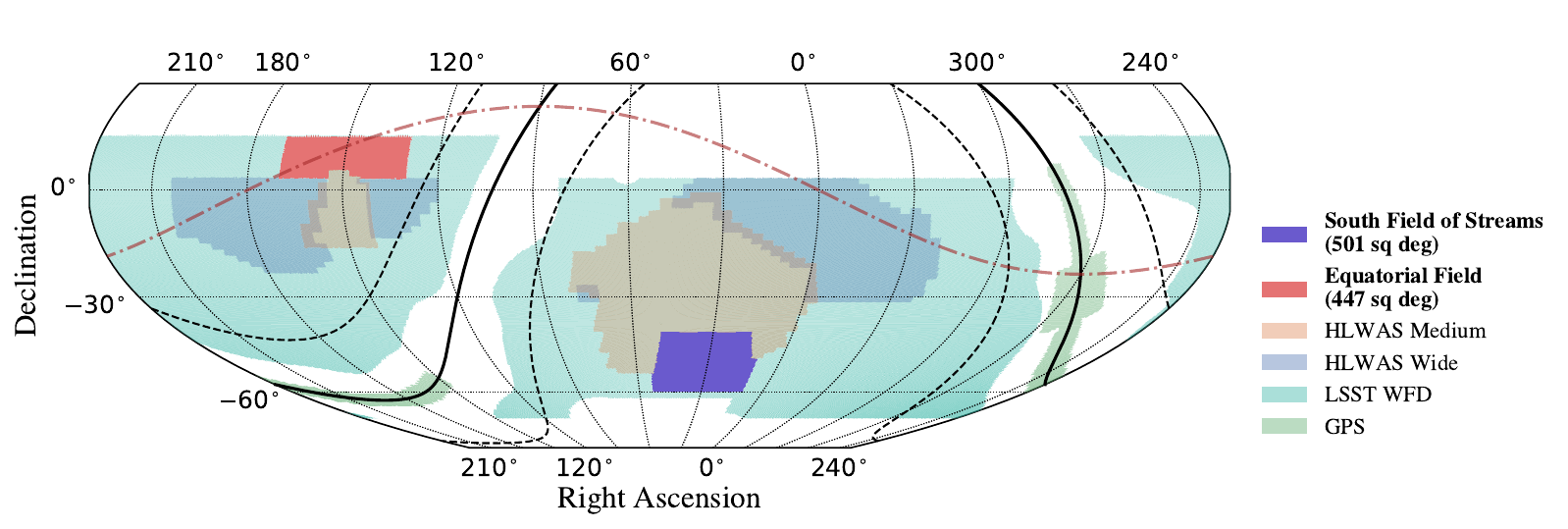}
    \caption{
    One potential Cycle 1 footprint (Equatorial Field + Southern Field of Streams), along with the coverage of other relevant surveys. These include the medium and wide components of the Roman High-Latitude Wide-Area Survey (HLWAS), the LSST wide-fast-deep (WFD) survey, and the Roman Galactic Plane Survey (GPS). The solid black line marks the Galactic plane, and the dashed black lines indicate $\pm20^{\circ}$ from the plane. The brown dash–dot line marks the ecliptic plane.
    }
    \label{fig:cycle1_footprint}
\end{figure}

In Figure \ref{fig:cycle1_footprint} we illustrate one possible Cycle 1 footprint that would triple the area of timely overlap between Roman Cycle 1 and LSST DR1. During Roman’s first two years, the majority of HLWAS observations are expected to be concentrated in the deep and medium tiers, covering a substantially smaller area than the wide tier. In this context, one possible Cycle 1 configuration would place the footprint adjacent to the medium tier, without repeating any wide-tier fields. Such an approach would increase the contiguous overlap between LSST DR1 and Roman Cycle 1 by $\sim1000,\mathrm{deg}^2$, roughly twice the projected medium-tier area at the time of LSST DR1. For field placement, two illustrative regions are highlighted: an equatorial strip, and a southern field containing prominent Milky Way substructures and stellar streams (i.e., a ``field of streams''). The equatorial strip could maximize early science return by enabling joint analyses across major Northern and Southern surveys, including DESI, PFS, LSST, Simons Observatory, 4MOST, and the Deep Synoptic Array, as well as all-sky surveys such as SPHEREx. We note that \textit{Euclid} does not cover this portion of the equatorial strip in order to avoid the ecliptic plane. In portions of this footprint with the highest zodiacal backgrounds (i.e., near the ecliptic plane), the achievable $H$-band depth would decrease to $\sim 25.0$, while only a few degrees away the depth would improve to $\sim 25.3$. The scientific return enabled by including this equatorial field may justify accepting a modest ($< 0.5$,mag) reduction in depth over a limited fraction of the footprint.

The southern field is optimized to encompass many known stellar streams and dwarf satellites, while also substantially increasing the likelihood of discovering new systems due to its proximity to the LMC. The number of stellar streams increases by a factor of $\sim2$ toward the LMC, as it contributes its own satellite population. In addition, the southern field renders the medium-tier footprint more circular, maximizing the number of galaxy pairs and thereby improving lensing statistics. Overall, the southern field given as an example can provide a rich resource for Galactic and extragalactic astronomy, as well as near-field and far-field cosmology. This wide early survey will facilitate the identification of rare objects, including quasars, rare galaxies, strong gravitational lenses, brown dwarfs, higher order stellar systems, and other serendipitous discoveries that will lay the framework for future wide-field imaging yields from Roman.

\subsection{Community Data Products}

Cycle 1 observations could support the development of several high-value data products for the community.

\begin{enumerate}

\item \textbf{LSST forced photometry at Roman source positions.}
One potential community effort would be to perform catalog-level LSST forced photometry at the positions of Roman detections across the proposed $\sim1000\,\mathrm{deg}^2$ of additional H-band imaging, as well as any HLWAS H-band coverage obtained during the first two years of Roman operations (anticipated to be $\sim500\,\mathrm{deg}^2$). Such processing could be carried out using the Rubin data pipelines and appropriate computational resources at S3DF.

\item \textbf{Joint \textit{Gaia}--Roman astrometric solutions.}
Another possible avenue would involve deriving joint astrometric solutions that improve proper motion measurements for faint \textit{Gaia} sources, for example using frameworks such as BP3M \citep{mckinnon24}. For $G\sim20$--21 stars, improvements of factors of $4$--$6$ relative to \textit{Gaia} DR4 may be achievable depending on cadence and depth.

\item \textbf{Catalogs of Galactic substructure and rare sources.}
Survey data could also support the construction of catalogs of Milky Way substructure (e.g., stellar streams, dwarf galaxies) and rare objects (e.g., strong gravitational lenses, brown dwarfs) identified within the footprint, facilitating rapid scientific exploration by the broader community.

\end{enumerate}

The community supporting the Roman all-sky survey brings together expertise in Rubin data processing, cross-mission astrometry, Galactic structure, and wide-field survey design, providing a strong foundation for the development of these data products in a timely and high-quality manner. We welcome broad participation and encourage contributions across these science efforts.

\section{Conclusions}

In this white paper, we have outlined a clear and executable path toward an all-sky survey with Roman that operates fully within nominal mission constraints while preserving substantial time for other ambitious programs. A Roman all-sky survey reaching $\mathrm{H}=25.5$ with $0.135''$ resolution would enable transformative science across cosmology, Galactic structure, stellar populations, and time-domain astrophysics, while establishing a legacy dataset for decades to come. For cosmology, such a survey would enable joint analysis of both Northern and Southern experiments (DESI, PFS, Simons Observatory, LSST etc), substantially increase the number of weak-lensing sources, improve the fidelity of photometric redshifts, and dramatically enhance star–galaxy separation and source deblending through joint analysis with Rubin. It would also expand the discovery space for strong gravitational lenses and deliver high-resolution galaxy imaging that can be studied in combination with \textit{Euclid}, SPHEREx, the Deep Synoptic Array, eROSITA, UVEX, and other large surveys spanning the electromagnetic spectrum. In the local universe, Roman can extend high-precision astrometry beyond the \textit{Gaia} magnitude limit, enabling proper motions and clean member selection for Milky Way satellites, stellar streams, and halo stars; mapping the outer halo and the LMC wake; and uncovering low-surface-brightness substructure across the sky. At the same time, an all-sky infrared reference image over 8 magnitudes deeper than 2MASS would provide foundational infrastructure for virtually every subfield of astronomy, enabling transient host identification, brown dwarf and wide stellar companion searches, imaging of globular cluster outskirts, and resolved stellar population studies in nearby galaxies.

Crucially, this science return does not require waiting for an extended mission. A first-epoch, single-band survey executed within the nominal five-year mission establishes the temporal baseline required for any future proper-motion or multi-band science, maximizes contemporaneous synergy with Rubin and JWST, and ensures that Roman’s discoveries can be rapidly pursued by the broader community. The Cycle 1 program described here illustrates one possible step toward such a long-term vision. In addition, this scenario will triple the area of timely overlap between Roman Cycle 1 observations and LSST DR1, ensuring a science legacy that is guaranteed irrespective of the outcomes of future observing cycles.

\section*{Acknowledgements}

This white paper was developed as an outcome of the workshop ``Proposing for the Roman Ultra-Wide Survey in the Era of LSST,'' held at Stanford University from January 26--28, 2026. The meeting was supported by an LSST-DA mini-grant and by the Kavli Institute for Particle Astrophysics and Cosmology. We thank the workshop participants for their contributions to the discussions, the online town hall attendees, and the collaborative development of the 
ideas summarized here.

\bibliography{references}{}
\bibliographystyle{aasjournalv7}



\end{document}